\documentclass[]{article}

\usepackage[utf8]{inputenc}
\usepackage{graphicx}
\usepackage{amsmath}
\usepackage{booktabs}
\usepackage{multirow}
\usepackage{longtable,booktabs}
\usepackage{graphicx}

\usepackage{url}

\title{Automated Classification of Airborne Laser Scanning Point Clouds}

\author{
Christoph Waldhauser \and Ronald Hochreiter
\and Johannes Otepka \and Norbert Pfeifer \and Sajid Ghuffar
\and Karolina Korzeniowska
\and Gerald Wagner
}

\date{April 2014}

\begin{document}
\maketitle

\begin{abstract}
Making sense of the physical world has always been at the core of
mapping. Up until recently, this has always dependent on using the human
eye. Using airborne lasers, it has become possible to quickly ``see''
more of the world in many more dimensions. The resulting enormous point
clouds serve as data sources for applications far beyond the original
mapping purposes ranging from flooding protection and forestry to threat
mitigation. In order to process these large quantities of data, novel
methods are required. In this contribution, we develop models to
automatically classify ground cover and soil types. Using the logic of
machine learning, we critically review the advantages of supervised and
unsupervised methods. Focusing on decision trees, we improve accuracy by
including beam vector components and using a genetic algorithm. We find
that our approach delivers consistently high quality classifications,
surpassing classical methods.
\end{abstract}

\section{Introduction}\label{introduction}

Surveying the very planet we live on has been an ongoing effort since
the dawn of mankind. From the early maps of Anatolia to modern
geospatial intelligence, the mission of any map was always to make sense
of the world around us. Boosting map drawing with the latest advances of
machine learning has the potential to largely facilitate the generation
of maps and extend their usefullness into application domains beyond
path finding.

In this chapter, we present the combined efforts of academia and
industry to create a framework for the automated generation of maps. The
basis for this project is airborne laser scanning: the systematic
recording and digitizing of ground by means of laser emitted from
aircraft. The resulting point clouds of the environment are then
automatically classified into ground cover types, using supervised
learning and evolutionary computation approaches.

This chapter is organized as follows. In a first section we describe the
technical background of airborne laser scanning. Section
\ref{classification} details the work related to develop automated
classification models. There we will compare the practical aspects of
supervised and unsupervised approaches as well as detail the supervised
classification approach we implemented and evolutionary computation
extensions to it. That section also features a description of the data
set we used to empirically test our approaches. The aspects of
implementing our model in industry applications is discussed in the
subsequent section \ref{industrial-applications}. We close with some
concluding remarks pointing to future research.

\section{Airborne Laser Scanning Point
Clouds}\label{airborne-laser-scanning-point-clouds}

\subsection{Measurement principle}\label{measurement-principle}

Airborne Laser Scanning (ALS) is a remote sensing method for obtaining
geometrical and additional information about objects not in contact with
the sensor, i.e.~the laser scanner. A laser scanner emits a short pulse
of infrared light which travels through the atmosphere and is scattered
and partially absorbed by any objects in the instantaneous field of view
of the laser beam. If diffuse reflection occurs, which is the standard
case for many object surfaces, including, e.g., vegetation, bare ground,
and building surfaces, a portion of the incident light is scattered back
to the sensor. There, the backscattered signal is detected and recorded.
The time lag between emission of the pulse and detection of its echo is
the two way travel time from the sensor to the object. With the known
speed of light this time lag is turned into the distance from sensor to
object. This is also called laser range finding (LRF). In laser
scanning, the beam is scanned across the entire field of view, thus
covering a larger extent . Rotating mirrors and comparable devices are
used to deflect the laser beam and cover large areas. With the known
orientation of the mirror and the known position of the laser scanner in
a global Earth fixed coordinate system (e.g.~WGS84, in UTM projection),
the location of the objects at which the laser pulse was scattered can
be computed. This provides a so-called 3D point cloud: a set of points,
each with 3 co-ordinates $x, y, z$. These points are obtained in the
sensor co-ordinate system.

\begin{figure}[t]{}
  \centering
  \includegraphics[scale=0.5]{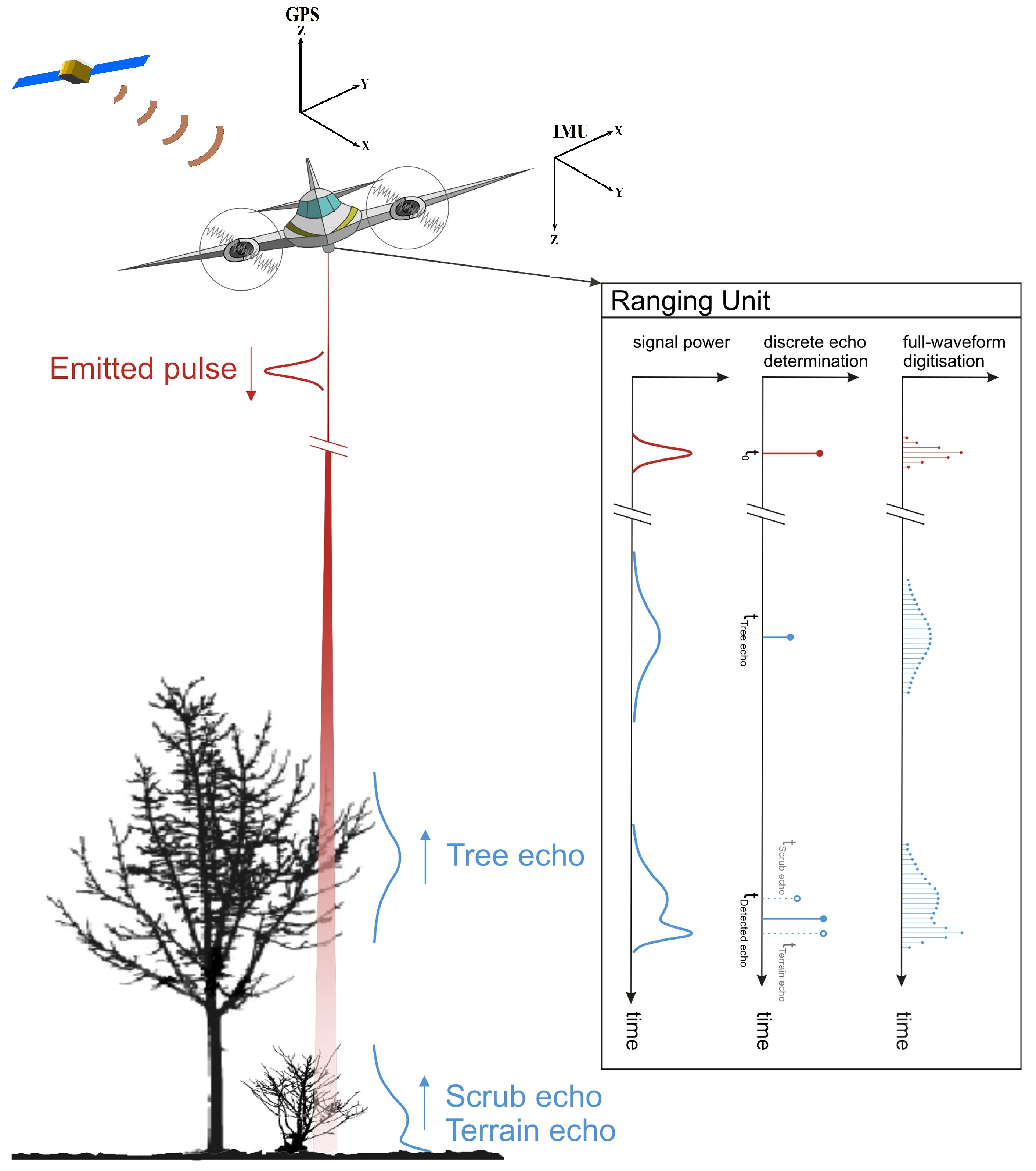}
  \caption{\label{fig:LS}Diagram explaining the principles of ALS \cite{Doneus2008}}
\end{figure}

In airborne laser scanning the scanner is mounted on a flying platform
(fixed wing or helicopter). Its position is measured with Global
Navigation Satellite Systems (GNSS, e.g., GPS). The angular attitude of
the sensor platform inside the aircraft is observed with Inertial
Measurement Unit (IMU, comprised of accelerometers and gyros). The laser
scanner is mounted to look downwards and the beam is scanned at right
angles to the flight direction (see Figure
\ref{fig:LS}\footnote{Full color, high resolution versions of each figure can be found at http://www2.wu.ac.at/alsopt}.).
Together with the forward motion of the aircraft, larger areas can be
scanned. Even larger areas are measured by flying strip-wise above the
terrain. This six degree of freedom trajectory defines a moving
co-ordinate system for the observation of range and angle from the laser
scanner. With an Euclidean transformation the points can be transformed
from the sensor co-ordinate system to the global co-ordinate system.
Typical results for the accuracy of such points is in the order of 10cm
(single standard deviation in each coordinate).

Besides the observation of distance between the sensor and an object
point by the time lag of emission and detection of its echo, also other
observations can be retrieved from the received echo. Firstly, it is not
always the case that the laser beam hits exactly one object. Due to the
diameter of the beam, e.g., 50cm, multiple objects may be within the
beam, but at different heights. Examples include vegetation canopy and
ground below. While a part of the signal is reflected at the leaves of
the canopy of a tree, other parts of the signal continue traveling
downwards until they hit lower vegetation or the ground, from which they
are reflected. Thus, each emitted pulse may give rise to several echoes.
Other examples, next to vegetation, are power lines and house edges,
where a part of the signal is reflected on the roof, while the other
part is reflected from the ground.

Furthermore, the backscattered echo can be sampled as a function of
time, so-called waveform digitizing. The recorded amplitude depends on
the range, on laser scanner device parameters, e.g.~the receiving
aperture diameter, but also on object properties, i.e., how much of the
incident signal is absorbed, scattered diffusely, etc. By means of
calibration \cite{Wagner2010ISPRS} the parameters of the object like the
backscatter cross-section, can be determined
\cite{WagnerEtAl2006,Roncat2011ISPRS}. The received echo may also be
deformed relative to the emitted pulse. An increased echo width
\cite{hoefle:silvi08} is a hint for either hitting a slanted surface or,
more often the case, hitting vegetation. Within the footprint a number
of leaves may be found which have similar by not identical height. Thus
the echoes from all the leaves overlap and form a single widened echo.

\subsection{Additional point
descriptors}\label{additional-point-descriptors}

A point cloud \textbf{P} is a collection of points
$p_i= \left(x,y,z\right) $ $\in$ \textbf{P} in a three dimensional
space. The laser scanning point cloud can be analysed locally to enhance
the description of each point further. For instance, given a point
density of 4~points/m$^2$, a local surface model
\cite{KrausPfeiferAnnapolis01} can be computed using e.g.~the ten
nearest points. This model may be an inclined plane, with it's normal
vector being an additional description for the point. The equation of a
plane through the point $\left(x_0,y_0,z_0\right)$ having a normal
vector $n=\left(a,b,c\right)$ is given as

\begin{equation}
a \left( x-x_0 \right) + b \left( y-y_0 \right) + c \left( z- z_0 \right) =0
\label{Plane}
\end{equation}

To compute the three components $\left(a,b,c\right)$ of the normal
vector, three equations, i.e.~three (non collinear) points are required.
To add robustness, generally more than three points are used. A subset
of points in the neighbourhood are typically selected based on k-nearest
neighbours or points with in the sphere of a pre-defined radius. If $k$
nearest neighbours are selected than there are $k+1$ points and
subsequently $k+1$ equations \eqref{Plane}. A least squares solution of
this overdetermined system of equations estimates an optimal plane by
minimizing squared sum of distances between the points and the estimated
plane. In the matrix form this equation system is written as

\begin{equation}
A \cdot \beta = 0,
\label{eqSystem}
\end{equation}

where each row of matrix $A$ contains the coordinates of a point
relative to the center $\left[x_n-x_0,y_n-y_0,z_n-z_0 \right]$, here
$n=1..k+1$ and $\beta=\left[ a,b,c\right]^T$ is the unknown normal
vector. The least squares solution for a system of equations of this
\eqref{eqSystem} form is equivalent to solving the eigenvalue problem of
the matrix $A^T A$. The unknown normal vector $\beta$ of the estimated
plane is the eigenvector corresponding to the smallest eigenvalue of
$A^T A$. The matrix $A^T A$ is often called structure tensor
\cite{gressin2012improving}. The mathematical form of the structure
tensor is:

\begin{equation}
A^TA=T=\frac{1}{k}\sum\limits_{i=1}^{k+1}  {\left( p_i - \bar{p} \right)}^T  \left( p_i - \bar{p} \right)  
\end{equation}

here $\bar{p} = \left(x_0,y_0,z_0\right)$ is the center of the points in
the neighborhood.

For house roofs or street surfaces the normal vectors have been shown to
reach an accuracy of a few degrees. The normal vector further allows to
convert the backscatter cross section into the so-called diffuse
reflectance. This value assumes a certain (Lambertian) scattering
behavior of the object. This scattering mechanism is described by the
reflectance (a unit-less value) and the normal vector of the surface. A
surface reflecting all incoming light perfectly diffuse has a
reflectance of 1.

The quality of the plane fitting, e.g., the root mean square distances
between the optimal plane and the given points indicates the roughness
of the surface \cite{Hollaus2011}. The smallest eigenvalue of $T$ gives
the variance of the distances between the points and the estimated
plane.

The structure tensor $T$ holds plenty more useful information about the
distribution of points in the neighborhood. The geometric information
encoded in $T$ is essential in the characterization and classification
of natural and artificial objects. Three widely used features derived
from $T$ are linearity, planarity and omnivariance. The linearity
feature reflects how well the distribution of points can be modeled by a
$3D$ line. Points over power lines exhibit such a characteristic,
therefore, the linearity feature is essential in classifying power lines
and similar structures. The planarity describes the smoothness of the
surface which is directly related to the roughness measure and the
quality of plane fitting for normal vector estimation. In contrast to
power lines and smooth surfaces, laser echoes from trees often spread
inhomogeneously across a larger $3D$ volume. This volumetric point
distribution is described by the concept of omnivariance. These features
are computed using the three eigen values
$\lambda_1 \geq \lambda_2 \geq \lambda_3 \geq 0$ of the matrix $T$:

\begin{equation}
L_T=\frac{\lambda_1-\lambda_2}{\lambda_1} 
\end{equation}

\begin{equation}
P_T=\frac{\lambda_2-\lambda_3}{\lambda_1} 
\end{equation}

\begin{equation}
O_T=\sqrt[3]{\lambda_1\lambda_2\lambda_3}
\end{equation}

In addition to $L_T$, $P_T$ and $O_T$, features like anisotropy,
eigenentropy and curvature are also derived using the eigenvalues of the
structure tensor $T$
\cite{mallet2008analysis,gross2006extraction,rusu2009semantic,weinmann2006extraction}.

More information about the characteristics of the surfaces can be
derived using features like \emph{echo ratio}, \emph{ZRange},
\emph{ZRank}, \emph{NormalizedZ} and \emph{PointDistance}. \emph{Echo
ratio} represents the vertical penetration of the surface
\cite{Hoefle2012134}. \emph{ZRange} represents the maximum height
difference between the points in the neighborhood, while \emph{ZRank} is
the rank of the point corresponding to its height in the neighboorhood.
\emph{NormalizedZ} is the rank of the point (between 0 and 1) multiplied
by the height range in the neighboorhood. \emph{PointDistance} is the
average of all shortest distances between the points in the
neigboorhood. A more detailed description of these features can be found
in \cite{OtepkaEtAl2013,OPALS}.

Thus, the point cloud can be augmented by additional parameters besides
the coordinates $x, y, z$: the echo ID (first, second, \ldots{} last
echo of a sequence of echoes) and overall length of the echo sequence,
echo amplitude, echo width, backscatter cross section, diffuse
reflectance, roughness (\emph{NormalSigma}), normal vector
(\emph{NormalX}, \emph{NormalY}, \emph{NormalZ}) echo ratio (\emph{ER}),
\emph{ZRange}, \emph{ZRank}, \emph{NormalizedZ}, \emph{PointDensity},
\emph{PointDistance}, linearity, planarity, and omnivariance.

\section{Classification}\label{classification}

A major application for the automated processing of point clouds is the
classification of points. In this application, every point is assigned a
class due to its inherent laser return characteristics and its derived
features. If successful, any such endeavor promises massive savings in
terms of human resources and time, and thus ultimately in cost.

In the past, the remote sensing community focused on classifying data
obtained from satellite measurements \cite{Mather2010}. They report that
results in general have been only somewhat satisfactory with large
portions being continuously misclassified. In contrast we work with air-
and not satellite-borne data. This allows for a much higher resolution
and considerable less atmospheric interference when measuring. Further,
the used full waveform data contains much more information than
traditional approaches using laser solely for range measurements.
Finally, the method of actively illuminating the ground with a laser
beam is superior to passively recording reflections of sun light.

Classification tasks can be grouped into human and machine based
classifications. Machine based classification itself can be split into
knowledge- and learning based systems. The former is today's industry
standard in ALS point cloud processing, the latter the eventual
developmental goal. The main disadvantage of knowledge-based systems
over machine learning classifiers is their requirement of explicit
definitions of ontologies and classification rules. Machine learning
classifiers, on the other hand base their classifications on rules
automatically deduced from the available data with minimal (or no) human
intervention. A machine learning classifier with human intervention uses
initial human input to deduce automatically classification rules from
it, that then can be used to autonomously classify points of previously
unseen point clouds.

When charging humans with point classification, a number of factors come
into play. Foremost, there is the need for additional data. Usually,
this data is provided by means of orthophotos that are (ideally) taken
in parallel to the laser scanning. Secondly, the qualification,
endurance and accuracy of the employed human has to be taken into
consideration as well. That person needs to be an expert user of
geographical information systems and trained to recognize the subtleties
of orthophotos.

This confluence of laser scanning data, external data via orthophotos
and human experience allows for rather precise classifications of
points. So far, human performance has not been surpassed by machines in
terms of accuracy. Naturally, human classification is a very time
consuming process. And equally naturally, machines outperform humans in
the time domain by many orders of magnitude. Therefore, investigating
algorithms for automated point cloud classification is an active area of
research.

When turning to learning based classification, two approaches following
the classical machine-learning dichotomy of supervised vs.~unsupervised
learning come to mind. The former requires initial human classifier
input to derive a classification of unseen points, while the latter does
not. The advantage of supervised classification is, that the resulting
classes correspond with target classes provided through human input.
Since unsupervised classification lacks any human interaction, the
classes found may or may not be interpretable or relateable to classes
that humans would come up with. In the remainder of this section, we
will focus on supervised classification based on initial human
interaction and the difficulties that arise from it.

As detailed above and elsewhere \cite{OtepkaEtAl2013}, point
characteristics can be grouped according to the way they were obtained:
by direct measurement, by calibrated or spatial improved measurements,
by deriving them computationally, by linking with meta data. For the
former three groups, problems can arise. Directly measured point
features are subject to the specifics of the laser scanner used. The
predominant method employed for airborne laser scanning enterprises is a
laser that is being deflected off the vertical by a rotating prism or a
swinging mirror. This allows to scan a range perpendicular to the flight
path and is essential for obtaining complete laser scans. However, this
method changes the characteristics of the laser return signal, as the
angle of the return signal not only depends on the characteristics of
the surface, but also the angle of the inbound signal. For instance,
when scanning directly below the aircraft only little occlusion will
occur, while at extreme scan angles, the laser beam will be obscured by
any objects between the aircraft and the ground. This distortion needs
to be taken into consideration when working with point cloud data.
Subsection \ref{border-effects} below discusses the detection of and
compensation for these effects in greater detail.

A further question that needs addressing is rooted in the way derived
attributes are being computed. Many such attributes are computed taking
in account a neighborhood of points. Here, neighborhood size becomes a
defining factor. Choosing an appropriate neighborhood size is far from
trivial. However, neighborhood size theoretically affects the
classification quality that can be derived. Further complications arise
from different neighborhood sizes that can be chosen for each attribute.
In subsection \ref{scale-space-selection} we present a genetic algorithm
for finding optimal neighborhood sizes for all neighborhood dependent
features involved in the classification.

Before turning to the problems described above, we will briefly
introduce the data set we worked with and describe how supervised
classification works from human and machine perspectives, respectively.

\subsection{Data set and example}\label{data-set-and-example}

From the industrial side of view, the motivation for this project was to
find a new, fast and reliable algorithm for the classification of point
clouds, which can minimize the manual checking and correction, because
every manual manipulation is a very time consuming task. The scenario
described in subsection \ref{flooding-prediction} below was the basis
for the development of the models used to automatically classify point
clouds.

The data set used was taken from the project \emph{DGM-W Niederrhein}
with kind permission of the \emph{Bundesanstalt für Gewässerkunde,
Germany}. Four predefined areas have been selected, each not bigger than
60 hectares, with different content like bridge, power lines, houses,
coniferous and deciduous trees, concrete, gravel, bare earth, groynes
and water.

The flight was done by airplane with the use of a Riegl LMS-Q560 200 KHz
Laser Scanner. Flight speed was 100 knots at an altitude above ground of
600 meters. The distance between the flight lines was 300 meters. The
effective scanning rate was set to~150 KHz with 80 lines per second. The
resulting mean point density was about 6 points/m$^2$ over the whole
area (except water areas). A radiometric calibration was computed using
asphalt streets in each flight session as calibration reference. The
calibration parameters were then applied to compute the reflectance, a
normalized intensity value. Roughness shapes (derived from digital
orthophotos), which define different ground classes of all areas were
known and used as support.

For the classification a list of classes was discussed and defined.
First a high-order list with standard classes (level 1), which are most
common in the majority of laser scanning projects was generated; then
each standard class was refined into subclasses (level 2) to better
represent the different kinds of environment. The classes used for this
project can be seen in Table \ref{tab:classes}.

\begin{table}
\centering
\caption{Defined classes in two levels of granularity.\label{tab:classes}}
\begin{tabular}{p{3cm}cp{4.5cm}c}
\toprule
\multicolumn{2}{c}{\textbf{First level}} & \multicolumn{2}{c}{\textbf{Second level}}\\
\cmidrule(r){1-2} \cmidrule(l){3-4}
Class & Code & Class & Code \\
\midrule
unclassified & 0 & unclassified & 0 \\
\midrule
undefined & 1 & undefined & 1 \\
\midrule
\multirow{7}{*}{ground} & & ground & 2 \\
 &  & sand & 18 \\
 &  & gravel & 3 \\
 &  & stone, rock & 4 \\
 &  & asphalt & 22 \\
 &  & cement & 21 \\
 & \multirow{-7}{*}{2} & river dam, groyne & 28 \\
\midrule
\multirow{3}{*}{vegetation} & & deciduous forest & 5 \\
 & & coniferous forest & 6 \\
 & \multirow{-3}{*}{5} & mixed forest & 7 \\
\midrule
\multirow{2}{*}{building} & & building roof & 8 \\
 & \multirow{-2}{*}{8} & wall, building wall & 24 \\
\midrule
water & 9 & water & 9 \\
\midrule 
 &  & car, other moving object & 10 \\
 &  & \noindent\parbox[c]{\hsize}{temporary object (under construction)} & 11 \\
 &  & bridge & 12 \\
 &  & power line & 13 \\
 &  & tower, power pole & 14 \\
 &  & bridge cable & 15 \\
 &  & road protection fence & 16 \\
\multirow{-9}{*}{artificial objects} & \multirow{-9}{*}{10} & bridge construction & 17 \\
\midrule
technical& 23 & \noindent\parbox[c]{\hsize}{technical, e.g. concrete part of a bridge} & 23 \\
\midrule
ground, vegetation & 20 & ground, vegetation & 20 \\
\midrule
error & 99 & error & 99 \\
\bottomrule
\end{tabular}
\end{table}

Following the refined class list and taking the roughness shapes into
consideration, a three dimensional classification was done manually
using the TerraScan software package. This was done to provide reference
data to generate training and testing data sets for the supervised
classification method. Later the manual classification was also used as
gold standard for assessing the results of the classification done with
both the supervised and unsupervised methods.

\subsection{Reference data generation through manual
classification}\label{reference-data-generation-through-manual-classification}

The term reference data refers to data that is manually classified by
humans using external data sources like orthophotos. It serves two
purposes: providing training data for supervised classification and
representing a gold standard that can be used to test the automatic
classification's accuracy.

One method to generate reference data is a manual classification of the
data set \cite{SitholeVosselman2004,KoblerEtAl2007}. This process
requires a thorough visual analysis of the data and a labeling of each
point. As this process is time consuming, typically only small parts of
the data set are manually labeled. Thus, this part should represent the
diversity of the terrain surface (flat, hilly, etc.) as well as a large
amount of the different target classes with a variety of geometric
appearance and distribution of other measured or derived objects. These
target classes often comprise natural objects (bare-earth, water,
vegetation, etc.) and man-made objects (buildings, roads, bridges,
ramps, power and other transmission lines, fences, cars and other moving
objects, etc.). Vegetation, as one targeted class for example, can be
tall and low and have different density. The variety of vegetation has
to be included in the manually labeled part for both, accuracy
assessment and machine learning. The diversity of classes depends on the
purpose of the classification.

Reference data generation can be performed in a number of different
ways. Firstly, an automatic classification based on a selection from
available algorithms can be performed, followed by a manual improvement
of the results. Secondly, only manual classification of an unclassified
data set can be performed. In the case of a large number of classes the
second way is recommended. A third option in the generation of reference
data by using existing data sets. As those data sets were often acquired
at a different time, with a different measurement technology, and often
with other applications in mind, the transferability of such a
classification is limited. Therefore, the next paragraphs will
concentrate on the methods for manual classification.

\begin{figure}
  \centering
  \includegraphics{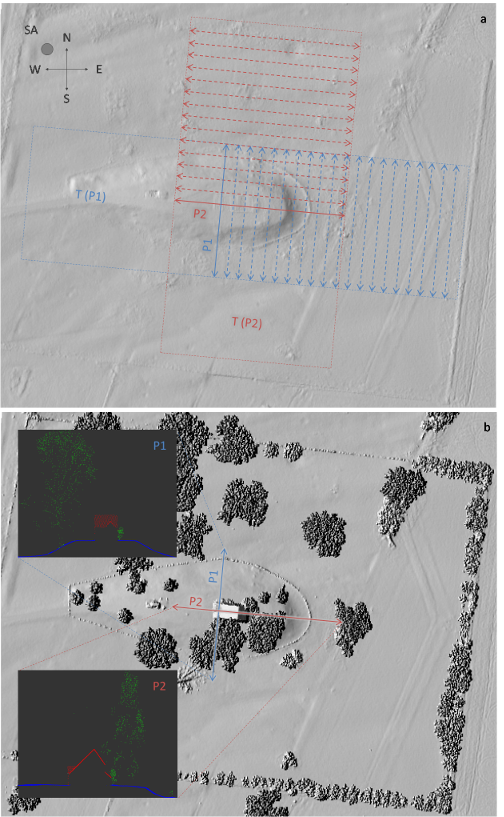}
  \caption{\label{fig:kk}Methods for visualizing the data during manual classification; a – shaded relief for DTM; b – shaded relief for DTM, buildings and vegetation; P1, P2 – 2D profiles; T(P1), T(P2) – transects for 2D profiles; SA – sun azimuth for shaded relief}
\end{figure}

The most common methods for visualization and reference data generation
are described below. The basic and most common method uses a 2D profile
(Figure \ref{fig:kk}). Profiles are sets of points cut out from the
entire point cloud with a vertical rectangular prism, not bound in
height. The width of the prism is typically small, e.g.~2 m, whereas its
length is larger, e.g.~50 m. These values are sensible when working with
point densities ranging from 1 to 20 points/m$^2$. Profiles allow the
user to see a part of the terrain from a side view which enables her to
distinguish the points within different classes but also to identify the
border between different objects, e.g.~building and ground, or
vegetation and ground. These borders are harder to identify in a top
view. In order to classify larger areas in an organized manner,
transects are used. This means that a set of parallel profiles is
generated which cover a rectangular area. Advancing in the manual
classification from one profile to the next accelerates the entire
process. The second method uses a shaded relief map (hillshade) of the
surface generated by the points of one class. A hillshade requires an
artificial illumination source, which is set in a standard manner to an
azimuth of 315 degrees, lighting the area from the northwest. Lighting
from different directions can substantially help to notice the terrain
slope as well as objects located on the ground, especially in the case
of mountainous regions. Hillshades can be generated for the bare-earth
class, in which the surface represents the digital terrain model (DTM).
An example for transects and hillshades can be found in the top panel of
Figure \ref{fig:kk}. That figure's bottom panel exhibits a DTM. Also
combinations of classes, e.g.~bare-earth and buildings, can be used.
This method can be applied for refining a manual classification,
i.e.~reclassifying points. This is especially suited to remove small
small artifacts which occur when close spatial proximity between two
classes led to a misclassification in an earlier step.

\subsection{Supervised classification}\label{supervised-classification}

The idea behind supervised classification is to automatically derive
from a small training set enough classification rules, so that a larger,
unseen data set can be classified automatically using the model derived
from the former. For that purpose, the training data needs to be
classified already. Usually, this initial classification is achieved by
manually classifying the points. This training data is then used to
build a model or equally train the classifying algorithm. In supervised
classification, the interpretation of the model comes second, therefore
more complex models are favored over simplistic ones that would ease
human interpretation; in fact, the boundary to model complexity is
dictated only by overfitting avoidance. This model is then used to
classify unseen data. To evaluate model performance, true classification
information for the unseen data is required as well. However, in
production environments, model evaluation for the entire data set is
usually not performed. Therefore, supervised classification promises to
save a considerable amount of costs.

The method of choice for supervised classification here is
classification trees. The tree is a predictive model that links up point
features with that point's class. Structurally, the tree consists of
leaves and branches. The leaves represent the final class labels and the
branches the conjunctions of features that lead up to these class
labels. Literature suggests a number of different algorithms for growing
a tree \cite{Safavian1991,Mather2010}. For the purpose of classifying
point clouds, we have found Breiman et al's Classification and
Regression Trees (CART) \cite{Breiman1984} to strike a good balance
between computational complexity and reliability. The implementation we
used was that of rpart \cite{rpart}. In terms of Friedl et al.
\cite{Friedl1997} these trees are univariate classification trees.

Conceptually, a classification tree seeks to partition the entire
feature space of a data set, one variable at a time. It does that by
selecting a variable and an appropriate splitting value that will
contribute maximally to node purity. Node purity is computed using the
Gini impurity coefficient:

\begin{equation}
I_G(f) = 1 - \sum_{i=1}^m f_i^2
\end{equation}

with $f_i$ being the fraction of items labeled to be of class $i$ for a
set of $m$ class labels.

This splitting and branch growing continues, until no variable can be
found that further increases node purity. The resulting trees can become
quite large which hinders interpretation (not a problem for point cloud
classification) and are prone to overfitting. This latter limitation can
become troublesome when trying to classify point clouds, as the learned
model does not generalize well anymore for unseen data. However, using
cross-validation and pruning off branches that are not occurring in a
significant number of replications, proves to be an effective tool
against overfitting.

As stated above, the performance of a classification tree can be gauged
if not only training but also test data contain true class labels. A
measurement statistic of classification performance is the
misclassification rate. Let $M$ be a cross-classification matrix between
true and predicted class labels and its elements being the counts of the
predicted elements and $J$ the number of all points in the point cloud,
then

\begin{equation}
MCR=1-tr(M)/J
\end{equation}

is the misclassification rate.

When selecting training data, two factors need consideration: the
randomness of the selection process and its stratification. The former
factor becomes important once large sets of random numbers need to be
created. While computers can always only generate pseudo random numbers,
most of them are sufficiently strong for point cloud
processing.\footnote{We used the R \cite{R-core} implementation of the
  Mersenne twister, which has a period of $2^{19937} - 1$.} However,
strong random number generation with guaranteed randomness does not
suffice to select a suitable training data set, if the classes are not
evenly distributed. In that common case, single classes---say temporary
construction structures---have only very few points associated with
them. When choosing points at random, it is extremely unlikely that many
of the rare class points will end up in the training data set. And if a
class does not show up in the training data set, the supervised
classification algorithm cannot learn the rules required to classify it.
Therefore simple random sampling schemes do not work in the presence of
rare classes.

To enable the supervised classification of rare classes, stratified
sampling needs to be applied. In its simplest form, stratified sampling
guarantees that numerous points from each class are selected for the
training data set. This, at the expense of having the entire training
data set being representative for the point cloud it has been sampled
from. The heuristic used for our stratified sampling approach sets the
size of the sample for stratum $c$ ($s_c$) to be either half of the
points of that class ($S_c$) or the overall sample size ($k$) divided by
the number of classes in the point cloud ($|A|$):

\begin{equation}
s_c=\mathrm{min}(\frac{S_c}{2}, \frac{k}{|A|})
\end{equation}

As noted above, the resulting stratified sample is not representative
for the entire point cloud anymore: rare classes occur much more often
in the training data set than they do in the point cloud. It is
therefore necessary to inform the supervised classification algorithm of
that misrepresentation.

\begin{figure}
  \centering
  \includegraphics[scale=0.5]{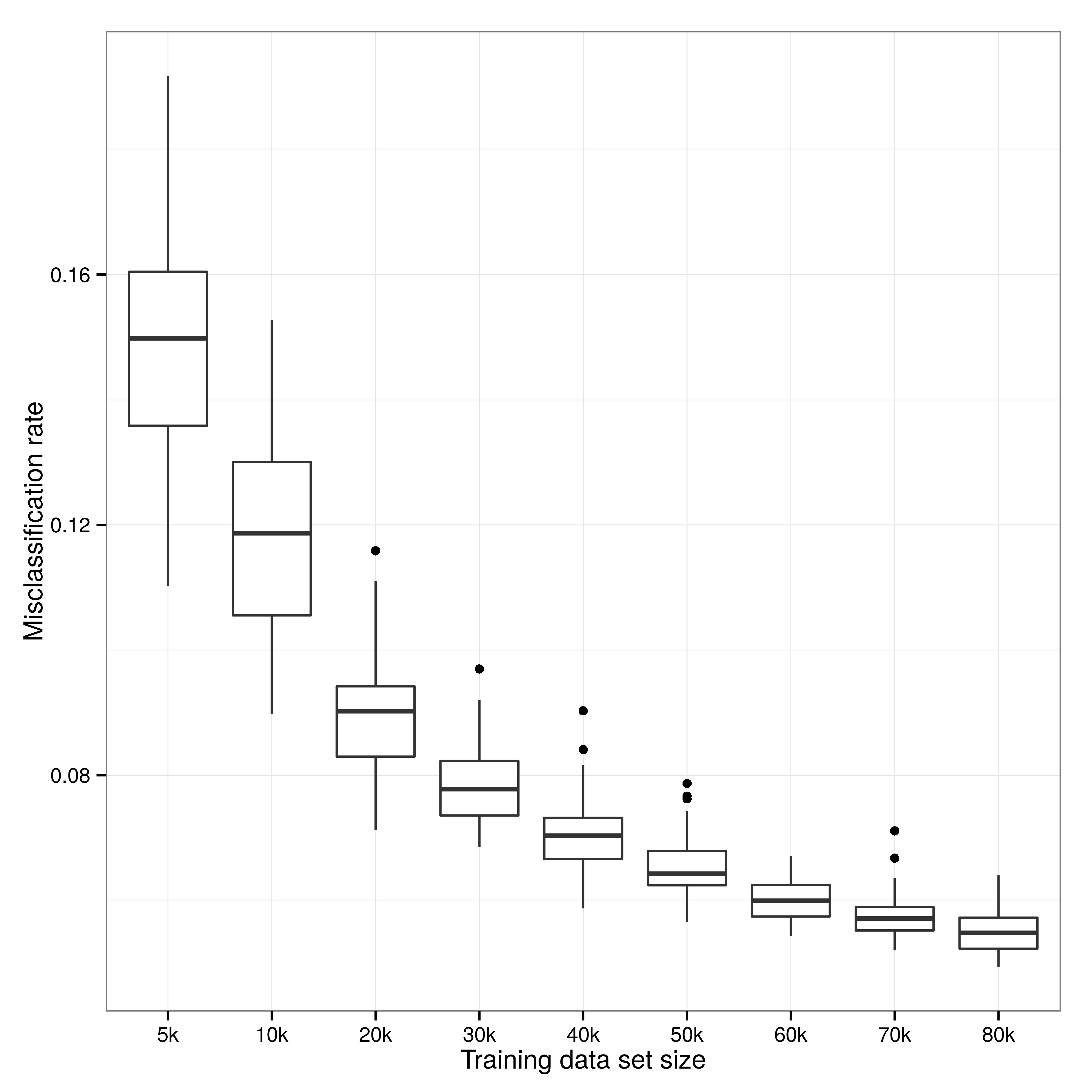}
  \caption{\label{fig:mcr-size}Misclassification rate as a function of training data size; classification of a 3 million strong point cloud, results bootstrapped with 50 replications}
\end{figure}

Perhaps obviously, the performance of a tree depends on the number of
data points it is allowed to learn from: the larger the training data
set, the better (usually) the classification of test data will be.
However, manually classifying points is expensive. Therefore, it is
crucial to find a training data set size that is just large enough to
produce reliable predictions. Figure \ref{fig:mcr-size} depicts this
relationship. As can be seen, there is a sharp drop between 10,000 and
20,000 points as training data set size with respect to mean
misclassification rate and its dispersion. After about 50,000 points,
the improvement gained by adding additional points subsides. We
therefore settled for 50,000 points as training data set size. The
resulting mean misclassification rate of 0.065 is a usable starting
point. In the following, we will discuss aspects of improving this
achievement even further.

\begin{figure}
  \centering
  \includegraphics[scale=0.25]{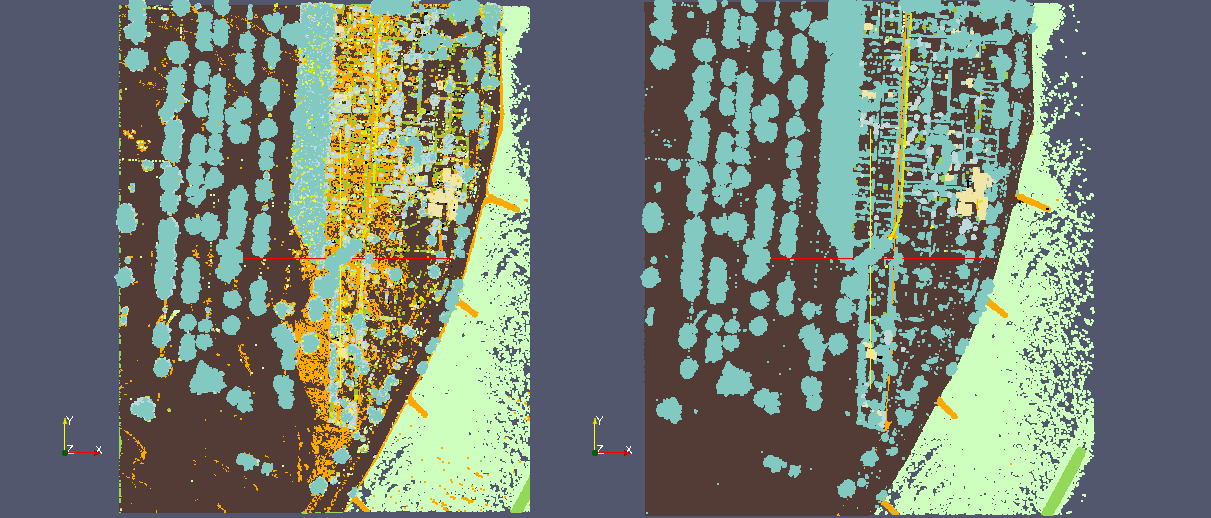}
  \caption{\label{fig:mc-graphic}Predicted (left) and true (right) classification of a sample area}
\end{figure}

When classifying point cloud data into predetermined classes, not all
classes that appear to be epistemologically justified to humans can be
sufficiently identified using laser return signals. For the problem at
hand, the points were to be partitioned into 26 classes. Logically,
these classes could be broken down into coarsely and finely grained
classes. While the coarse classes were successfully classified (MCR:
0.02, $\sigma=0.002$), the finer classification exhibited the 6.5\% MCR
as described above. Table \ref{tab:prob-classes} lists the finely
grained classes that were notoriously troublesome. Figure
\ref{fig:mc-graphic} shows the differences between automatic and true
human classification results.

\begin{table}
\centering
\caption{\label{tab:prob-classes} Classes that were hard to predict. Percentage of points that ended up in that class. Remainder to 100 percent is scatter in all classes.}
\begin{tabular}{ll}
\toprule
\textbf{True class} & \textbf{Predicted classes}\\
\midrule
\multirow{3}{*}{Building, wall} & Deciduous~forest~(67\%)\\
 & Building~roof~(17\%)\\
 & Building,~wall~(17\%)\\
\midrule
\multirow{2}{*}{Temporary~object} & Temporary~object~(78\%)\\
 & Road~protection~fence~(14\%)\\
 \midrule
\multirow{2}{*}{Power~pole} & Power~pole~(75\%)\\
 & Road~protection~fence~(16\%)\\
\midrule
Error class points & Scattered in all classes\\
\bottomrule
\end{tabular}
\end{table}

When casting a more detailed look at these misclassifications, it
becomes evident that many of them are conceivably caused by imprecise
classifications of humans in the first place. Consider for example a
road in winter: the asphalt tarmac is at places covered with grit sand
to prevent the icing of the road. Grit and asphalt tarmac will differ in
texture and material. Therefore, the laser return signal for patches of
road that contain more grit sand than others will exhibit different
characteristics. In manual classification based on aerial photography,
these patches of grit sand are unlikely to be identified and marked as
such by the human classifier. To a certain extent, the misclassification
rate achieved by supervised classification of finely grained classes can
be explained by the algorithm outperforming human classification. This
is obviously very dependent on the quality of human classification.

A similar argument holds for the error class. Here, points were
classified as errors if some of their measurements exceeded a valid
measurement range. The algorithm was informed about these missing
values. On the other hand, classification trees are able to cope with
missing information by substituting it with the second best split.
Therefore, points that a human would not classify because it contained
obviously faulty measurements, were classified by the algorithm.

Another problem that is rooted in the difficulty of epistemological
concepts is the misclassification of many temporary object points as
road protection fences. It is difficult for any automated classifier to
learn the concept of an object being temporary in nature. While the
algorithm successfully classifies almost all temporary objects as some
kind of artificial objects, it cannot differentiate between these
objects being permanent or temporary (road protection fences).

\begin{figure}
  \centering
  \includegraphics[scale=0.18]{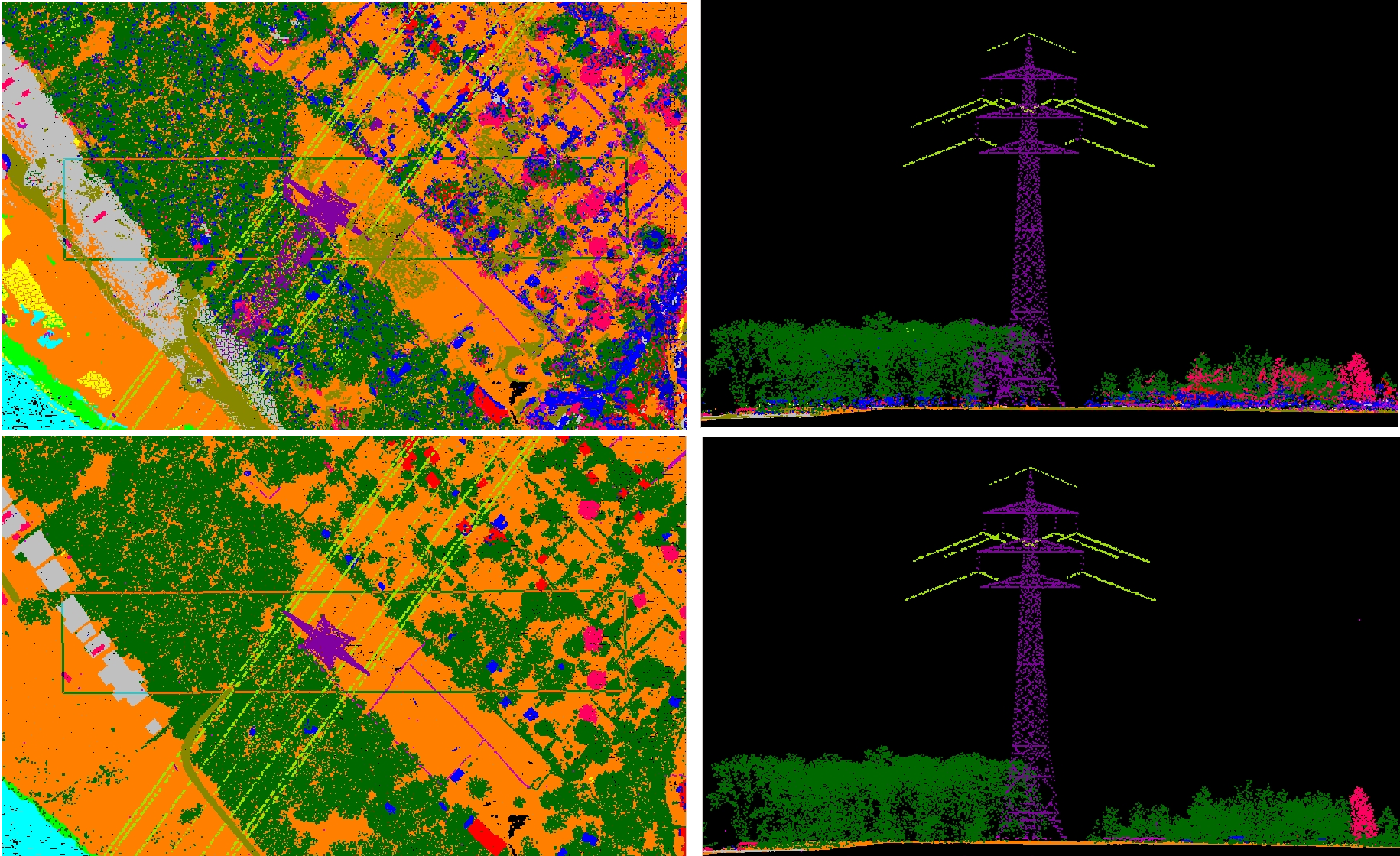}
  \caption{\label{fig:mc1}Misclassification at a power line where pole and vegetation cannot be separated reliably; automatically classified point cloud on top, bottom panel shows the manually classified one}
\end{figure}

However, the largest problem in misclassification cannot possibly be
rooted in epistemological complexities: Buildings and walls are being
classified predominantly as trees. From a geometrical point of view,
trees and buildings do indeed share some properties related to their
height and volume. On the other hand, distinctive characteristics like
texture and material should have been picked up by the algorithm. This
type of mistake is also represented in Figure \ref{fig:mc1}. To some
extent, the misclassifications can be explained by snow or leaf covered
roofs on top of buildings. Still, this unsatisfactory performance can
most likely only be overcome by implementing geometrical shape detection
in a post-processing step. This is the focus of ongoing research.

\subsection{Border effects}\label{border-effects}

Airborne laser scanning is limited by the principles of optics:
dependent on the incident angle, the characteristics of a laser return
signal varies. For example, hitting vegetation from the side will
produce many more laser echoes than hitting it straight from above.
Also, the shape of the beam's cross section depends on that angle.
Additional distortion in the characteristics of points may arise from
the method of aerial laser scanning. Due to the limited field of view of
airborne laser scanners wider areas are scanned by multiple overlapping
strips. Typically, these strips overlap to achieve full coverage even in
case of wind sheer or minor navigation errors. In these overlapping
areas, the properties of the measurement process change (as there are
multiple overpasses); a change that needs to be accounted for.

One method to compensate for the different return signal
quality/properties is to take the deflection of the laser into account.
There are two approaches available. One uses the raw beam vector
components ($v_x$, $v_y$, $v_z$) that indicate the deflection of the
laser beam for a given point. The other method combines these components
to derive the scan angle $\phi$:

\[\phi = \mathrm{arctan}(\frac{\sqrt{v_x^2+v_y^2}}{|v_z|})\]

The following Table \ref{tab:be} shows the effect beam vector components
and scan angle have on the misclassification rate. Starting with the
simplest model without any compensation for border effects, the mean
classification rate lies at 8.1\%. Adding the scan angle to the model
improves its quality by one, beam vector components by two percentage
points. Adding both compensation terms to the model barely improves
classification quality with respect to a pure beam vector components
model.

\begin{table}
\centering
\caption{\label{tab:be} Model quality in mean MCR for models with different kinds of border effect components. Results bootstrapped with 50 replications.}
\begin{tabular}{p{4cm}rr}
\toprule
\textbf{Model type} & $\boldsymbol{\mu_{MCR}}$ & $\boldsymbol{\sigma_{MCR}}$ \\
\midrule
no border effects & 0.081 & 0.004 \\
beam vector components & 0.065 & 0.005 \\
scan angle & 0.074 & 0.005 \\
\noindent\parbox[c]{\hsize}{beam vector components and scan angle} & 0.063 & 0.004 \\
\bottomrule
\end{tabular}
\end{table}

\subsection{Scale space selection}\label{scale-space-selection}

A number of point cloud features are not directly measured but computed
with respect to any points immediate neighborhood. In general, the local
neighborhood of a point can be defined in 2D or 3D. Furthermore, a
certain number of closest neighbors, a fixed distance or a combination
of both can be used as neighborhood definition. For the following
analysis a cylinder (i.e.~2D fixed distance neighborhood) for each point
is formed. Obviously, larger radii lead to a stronger averaging effect
while smaller ones are prone to overfitting. It, therefore, is important
to find the optimal radius for each feature in order to minimize
misclassification rate.

To discover the optimal radii for neighborhood-dependent features, a
genetic search algorithm \cite{Goldberg1988} was used. In the following
we will describe the genetic algorithm used for this optimization and
its parameters. We then turn our attention towards evaluating the
algorithm's performance in terms of convergence and solution stability.
The former examines the relation of improvement achieved due to and time
spent on optimization. The latter analyzes the stability of recommended
radii across a number of optimizations.

The 13 neighborhood-dependent features were computed each with radii
ranging from 1 to 6~m in 0.5~m increments resulting in 11 versions of
each feature. The algorithm's genomes were then modeled to be integer
vectors of length 13 with each gene being an integer from 1 to 11,
encoding the chosen neighborhood size for each feature. The algorithm
was initialized with 100 random genomes as starting solutions. The
standard genetic operators of single-point cross-over breeding and
mutation were employed for evolutionary optimization. Further, pairing
genomes for mating was done using tournament selection and a proportion
of the top performing solutions was cloned directly into each new
generation. To ensure that the gene pool remained fresh and to safeguard
against local optima traps, some random genomes were introduced with
each generation. Table \ref{tab:ga-props} gives the parameters of the
genetic algorithm, which were established by experiment.

\begin{table}
\centering
\caption{\label{tab:ga-props} Parameters of the genetic algorithm.}
\begin{tabular}{lr}
\toprule
\textbf{Parameter} & \textbf{Value} \\
\midrule
Population size & 100 \\
Tournament size & 5 \\
Mutation probability & 0.05 \\
Elite proportion & 0.1 \\
Reseed proportion & 0.1 \\
\bottomrule
\end{tabular}
\end{table}

The fitness function to be optimized was the misclassification rate as
described above. In order to ensure comparability, MCR was computed
using the same training--test data split each time. The initial split
was generated using a stratified sampling scheme and included 5137
points in the training data set. Using a random sample of 100,000
points, the algorithm was allowed 500 generations to find the optimum
combination of radii for the 13 neighborhood-dependent features. In
order to ensure computability within reasonable time, not the entire
point cloud could be processed. Therefore, a very large simple random
sample of 100,000 points was drawn from the point cloud, and all
operations were performed on that sample.

As genetic optimization is essentially stochastic in nature, the
optimization was repeated 34 times. Of these 34 replications, 30 reached
the same optimum while 4 stayed behind (by a very small margin). Almost
all replications had converged to the optimum after 50 generations. By
generation 75 all 30 successful replications had converged. The optimum
discovered implied a misclassification rate of 0.022. When compared to
the best misclassification achieved using a constant radius of 6 meters
(0.065) this is a notable improvement by more than 60 percent.

\begin{figure}
  \centering
  \includegraphics[scale=0.7]{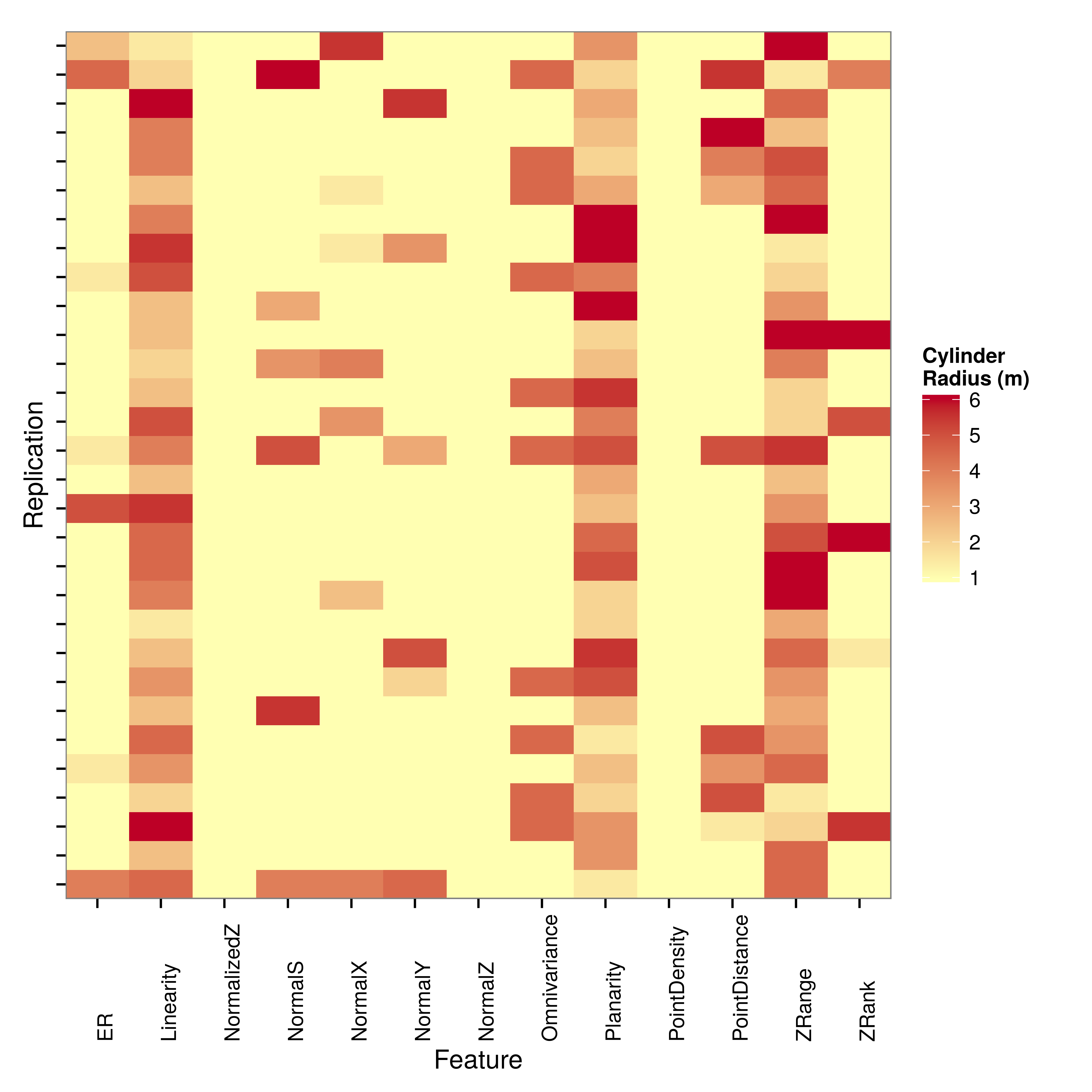}
  \caption{\label{fig:ga-stab}Solution stability of 30 genetic optimization replications}
\end{figure}

Turning to solution stability, it is of interest whether each
replication's terminal solution leads to the same combination of radii
or not. Figure \ref{fig:ga-stab} displays a heat map of cylinder radii
per feature chosen in each (optimal) replication. Features that exhibit
the same color shades for the entire column can be considered stable.
These are the variables NormalizedZ, NormalZ and PointDensity. For each
of these features, the optimal cylinder radius is at 1 meter. At the
other end of the spectrum, very colorful columns, Linearity, Planarity
and Z-Range, are indicative of features whose neighborhood size has no
impact on misclassification rate.

The genetic algorithm delivers a definite improvement of the
misclassification rate. The remaining two percent are most likely due to
measurement and human classification error. With respect to solution
stability, it became obvious that while some features are
computationally dependent on neighborhood size, the outcome is not
affected by them. On the other hand, there are features that clearly
exhibit a strong dependence on neighborhood size. Conceptually, the
genetic algorithm can be improved by implementing consensus voting when
delivering radii recommendations. This too, is an ongoing research
effort.

We conclude that supervised classification of point clouds is definitely
an idea worthwhile pursuing. The data quality obtained from airborne
laser scanning allows for a very precise analysis of the ground. In
combination with the sophisticated computation of derived point cloud
features, advanced classification algorithms sampling schemes as well as
evolutionary optimization strategies, we are able to produce
classification accuracies that surpass classical satellite based
classification. While the classical approaches rarely ever reach above
90 percent accuracy, our approach delivers consistently accuracies close
to 100 percent. While there are challenges that remain to be overcome,
the achieved accuracy is already good enough for many applications. In
the following we will discuss these applications further.

\section{Industrial applications}\label{industrial-applications}

Airborne Laser Scanning is in use for industrial purposes since the mid
1990s and has dramatically improved since then. For expample: in the
beginning there have been laser scanners with a fixed array of fibre
optical conductors, which brought a good point density in the direction
of flight, but very poor density in the transverse direction. So a
detection of embankments along the flight direction was very hard.
Technological advances like the steadily increased measurement rates,
improved apertures and new detection algorithms prepared the way for a
wide field of applications.

There are different technologies at work in today's laser scanners: they
provide sampling rates of up to 600,000 laser pulses per second. Also
modern apertures are able to detect more than just one single return per
pulse and provide reflectance, echo ID and echo width for each return;
some can even penetrate water surfaces and give information on submarine
ground and submerged objects.

Higher point densities result in better environment depicting. With
today's high point densities, embankments can be well detected by
extracting \emph{breaklines} within the point clouds. Normally 4
points/$m^2$ will be ordered, but customers more often want 8 or more
points/$m^2$. This gives the opportunity to model the ground more
precisely. But customers are not only interested in the presence of
ground, they also want to know what kind of ground they are looking at.

Classification is mostly a semiautomatic process, consisting of an
automatic step and a manual checking and correction step. One of the
aims is to minimize the need of manual correction, due to its cost.
Another aim is to improve the automatic detection of more than a
standard set of classes to cater to future customer's requirements.

In the following we will present some examples of airborne laser
scanning applications.

\subsection{Digital Terrain Model}\label{digital-terrain-model}

Often a plain model of the ground is needed for planning or research
purposes. These models are of great importance for e.g.~road- or railway
planning offices, to know how much material has to be removed or added
for street or railway planning. Therefore the point cloud has to be
classified with special emphasis on detecting erroneous echoes. The DTM
classes mostly consist of ground, water and unclassified points, which
have no influence on the model.

\subsection{Digital Surface Model}\label{digital-surface-model}

The DSM features ground, vegetation, buildings, bridges and sometimes
power lines and describes the earth's surface including natural and
artificial objects. By subtracting the DTM from the DSM the result will
be a normalized DSM. This can then be used for e.g.~easy measurement of
building or vegetation heights.

\subsection{Avalanche prediction}\label{avalanche-prediction}

In mountainous areas avalanches (snow or boulders) are a common threat,
so prediction and subsequently protection is an important task. For
aviation purposes it is also necessary to know the position of power
lines or cable-cars. Therefore each point needs to be classified along
the lines of ground, various vegetation, water, building, power lines,
\ldots

To compute the pathways and probabilties of avalanches in certain areas,
one not only needs to know point classes, but also inclination,
roughness (in this case roughness refers to a parameter, which will tell
how fluids will be slowed on a surface), azimuth, \ldots

All these features can be derived out of the point cloud by classifying
using the above algorithm.

\subsection{Flooding prediction}\label{flooding-prediction}

To protect people and environment in areas that are in danger of
flooding around rivers, it is vital to know, how water is flowing over
different types of ground. Therefore ground has to be classified in
different roughness classes, that have known properties for flowing or
seeping. The classification of roughness areas is normally done by
digitizing digital orthophotos \cite{WSV1,WSV2}. In respect to the
classification methods described in section \ref{classification},
roughness can be set in direct relation with different ground classes.
Taking into account the derived DTM together with the digitized
breaklines \cite{TUW-177260}, a triangulated surface can be computed.

By combining the DTM surface with information of the different point
classes from ground detection, there can be defined areas with varying
roughness. This classification is normally done by using digital
orthophotos as reference. By classifying the roughness purely from the
data contained within a laser point cloud, the high cost of extra
orthophotos can be skipped.

\subsection{Forestry and Vegetation}\label{forestry-and-vegetation}

The detection of forested areas is an important part of environmental
applications. Especially time series analyses, e.g.~to estimate
deforestation, was often carried out using analog or digital orthophotos
so far. However, Airborne Laser Scanning gets more popular for such
applications, because it is not restricted to the canopy. The laser beam
can often penetrate the vegetation returning multiple echoes. This
provides information about the vertical structure of the forest
including good knowledge of the ground, which is needed to compute high
quality DTMs, tree heights, stem volumes, etc. In urban areas the
knowledge of classified vegetation is used in applications for 3D
visualizations, urban planning, noise emission charts, etc.
\cite{rutzinger2007detection}.

\section{Conclusion}\label{conclusion}

In this chapter we presented an overview of advances in processing and
automatically classifying point clouds from airborne laser scanning.
Particularly, the accuracy of the classification of point clouds can be
improved greatly using machine learning based methods like decision
trees. There, manually classified training data---a small subset of the
entire point cloud---is used to build a classification model. This then
in turn can be used to classify the remainder of the point cloud or a
fresh one.

These advances in classification accuracy are chiefly due to our making
use of the entire full wave form of the laser echoes. Using advanced
radiometric and computational methods, for every echo additional
properties or features are computed from that echo's wave form, external
data and the echo's immediate neighborhood. Using an evolutionary
algorithm we were able to identify features where the size of that
neighborhood influenced classification accuracy and establish optimal
neighborhood size values for these features.

The model presented in this chapter has applications ranging from
forestry to avalanche and flooding protection. A more immediate
application is the automatic generation of maps. However, this is but
the beginning of our journey. We already pointed to the inclusion of
shape detection for improving classification accuracy and consensus
voting the genetic algorithm to optimize neighborhood size
recommendations as current research goals. Further extensions focus on
better understanding how the scan angle affects echo properties when
analyzing the flights strip-wise. A major issue is the possibility to
learn from multiple but possibly unreliably sources. Often, orthophotos
related to a point cloud are out-of-date or older maps are used to
provide external reference data. Ideally, if we were able to use these
data sources to speed up training data and model generation, the entire
remote sensing work flow could be revolutionized.

\bibliographystyle{spmpsci}

\bibliography{alsopt}

\end{document}